\pdfoutput=1

\documentclass[11pt]{article}

\usepackage[final]{acl}

\usepackage{times}
\usepackage{latexsym}

\usepackage[T1]{fontenc}

\usepackage[utf8]{inputenc}

\usepackage{microtype}

\usepackage{inconsolata}

\usepackage{hyperref}
\usepackage{url}

\usepackage{booktabs}       
\usepackage{amsfonts}       
\usepackage{nicefrac}       
\usepackage{xcolor}         
\usepackage{graphicx} 
\usepackage{amsmath}
\usepackage{adjustbox}
\usepackage{multirow}
\usepackage{enumitem}

\title{ControlSpeech: Towards Simultaneous and Independent Zero-shot Speaker Cloning and Zero-shot Language Style Control}

\author{
 \textbf{Shengpeng Ji\textsuperscript{1,2}},
 \textbf{Qian Chen\textsuperscript{2}},
 \textbf{Wen Wang\textsuperscript{2}},
 \textbf{Jialong Zuo\textsuperscript{1}},
\textbf{Minghui Fang\textsuperscript{1}}, \\
 \textbf{Ziyue Jiang\textsuperscript{1}},
 \textbf{Hai Huang\textsuperscript{1}},
 \textbf{Zehan Wang\textsuperscript{1}},
 \textbf{Xize Cheng\textsuperscript{1}},
 \textbf{Siqi Zheng\textsuperscript{2}},
 \textbf{Zhou Zhao\textsuperscript{1}\thanks{Corresponding author.}}
\\
 \textsuperscript{1}Zhejiang University~~~~
 \textsuperscript{2}Alibaba Tongyi Speech Lab
}

\begin{document}
\maketitle
\begin{abstract}
In this paper, we present \textbf{ControlSpeech}, a text-to-speech (TTS) system capable of fully cloning the speaker's voice and enabling arbitrary control and adjustment of speaking style. Prior zero-shot TTS models only mimic the speaker's voice without further control and adjustment capabilities while prior controllable TTS models cannot perform speaker-specific voice generation. Therefore, ControlSpeech focuses on a more challenging task—a TTS system with controllable timbre, content, and style at the same time. ControlSpeech takes speech prompts, content prompts, and style prompts as inputs and utilizes bidirectional attention and mask-based parallel decoding to capture codec representations corresponding to timbre, content, and style in a discrete decoupling codec space. Moreover, we analyze the many-to-many issue in textual style control and propose the \textbf{Style Mixture Semantic Density (SMSD) module}, which is based on Gaussian mixture density networks, to resolve this problem. To facilitate empirical validations, we make available a new style controllable dataset called \textbf{VccmDataset}. Our experimental results demonstrate that ControlSpeech exhibits comparable or state-of-the-art (SOTA) performance in terms of controllability, timbre similarity, audio quality, robustness, and generalizability. Codes are available at \url{https://github.com/jishengpeng/ControlSpeech}.
\end{abstract}

\begin{figure}[htbp]
\centering
\includegraphics[height=3cm, width=7.6cm]{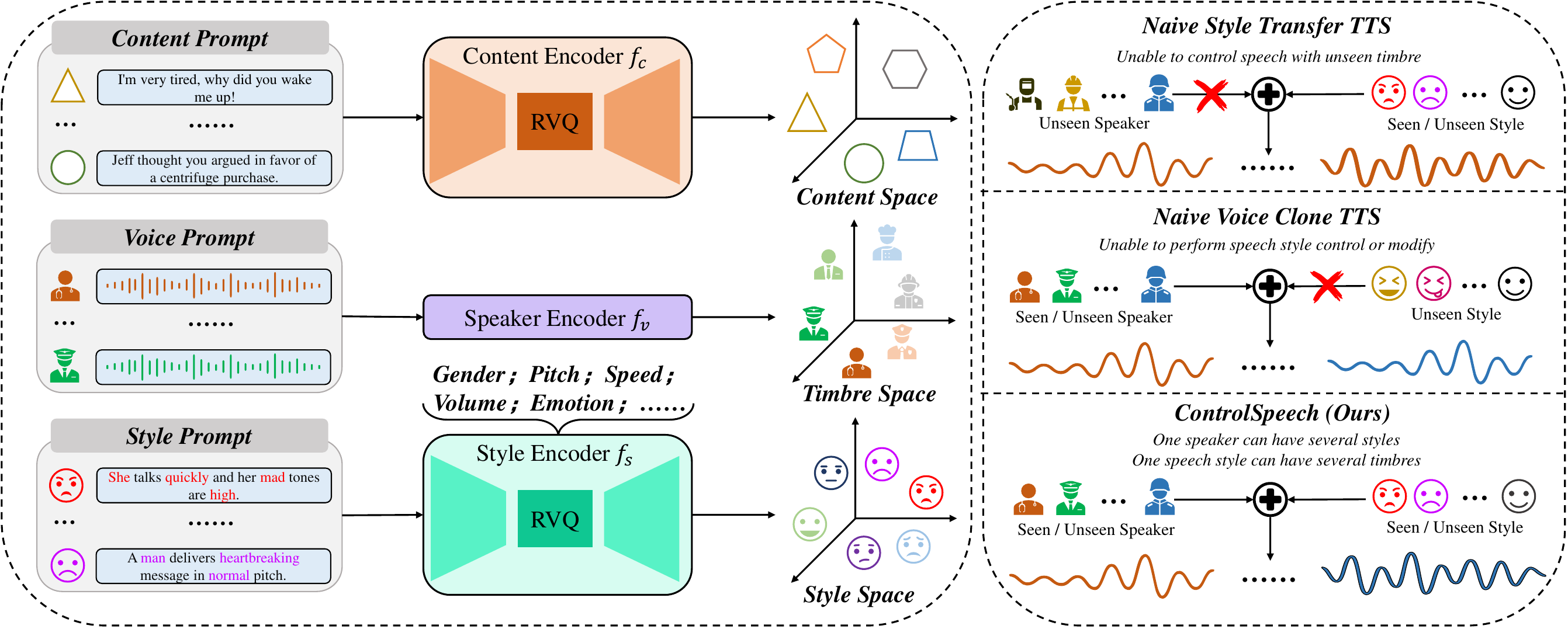}
\caption{The voice prompt, the content description, and the style description correspond to the timbre, content, and style representations in the discrete codec space in the left panel . The right panel compares ControlSpeech with previous style-controllable TTS and zero-shot TTS systems. In this comparison, we use the amplitude and color of the waveform to represent the styleand timbre.}
\label{arc1}
\end{figure}

\section{Introduction}
\label{sec:intro}
Over the past decade, the field of speech synthesis has seen remarkable advancements~\citep{fastspeech2,vits}, achieving synthesized speech that rivals real human speech in terms of expressiveness and naturalness~\citep{naturalspeech}. Recently, with the development of large language models~\citep{gpt,touvron2023llama} and generative models in other domains~\citep{ho2020denoising,flow1}, the tasks of zero-shot TTS~\citep{valle,naturalspeech2,le2023voicebox} and style-controllable speech synthesis~\citep{prompttts,instructtts} have garnered significant attention in the speech domain due to their powerful zero-shot generation and controllability capabilities. Zero-shot TTS~\citep{speartts} refers to the ability to perfectly clone an unseen speaker's voice using only a few seconds of a speech prompt, commonly achieved by significantly scaling up both the training data and model sizes. On the other hand, style-controllable TTS~\citep{prompttts} supports the control of a speaker's style (prosody, accent, emotion, etc.) through textual descriptions.

However, these two types of models have their own limitations. As illustrated in the right panel of Figure~\ref{arc1}, prior zero-shot TTS \citep{valle} can clone the voice of any speaker, but the style is fixed and cannot be further controlled or adjusted. Conversely, prior style-controllable TTS~\citep{prompttts2} can synthesize speech in any desired style, but it cannot specify the timbre of the synthesized voice. Although some efforts~\citep{instructtts,promptstyle} have been made to use speaker IDs to control the timbre, these approaches are limited to testing on constrained in-domain datasets and lack the zero shot ability. As a result, current speech synthesis systems lack \textbf{independent} and flexible control over \textbf{content, timbre, and style at the same time}, for example, they are unable to synthesize speech in Trump's voice with a child's joyful style saying ``Today is Monday''. To address these limitations, we propose a novel model called ControlSpeech. To the best of our knowledge, ControlSpeech is the first model to \textit{simultaneously} and \textit{independently} control timbre, content, and style, and demonstrate competitive zero-shot voice cloning and style control abilities.

There are two main challenges to achieve simultaneous and independent control over content, timbre, and style in a TTS system. First, the information from the style prompt and the speech prompt can become entangled and interfere with or contradict each other. For instance, the speech prompt might contain a style different from that described by the textual style prompt; therefore,  simply adding a style prompt control module or a speech prompt control module to previous model frameworks~\citep{prompttts2,valle} is evidently insufficient.  Second, there lacks large datasets that fulfill both requirements of zero-shot TTS systems and textual style-controllable TTS systems. Specifically, due to the scarcity of style-descriptive textual data, the training data for mainstream style-controllable TTS systems~\citep{prompttts,promptstyle} typically amounts to only a few hundred hours~\citep{textrolspeech}, far from meeting the requirements of a large-scale, multi-speaker training dataset~\citep{librilight} that is crucial to attain robust zero-shot speaker cloning capabilities.  To tackle these two challenges, we explore a novel approach in ControlSpeech that leverages a pre-trained disentangled representation space
for controllable speech generation. On one hand, disentangling representations enables independent control over content, style, and timbre. On the other hand, utilizing a representation space pre-trained on a large-scale multi-speaker dataset
ensures robust zero-shot capabilities of ControlSpeech. In this work, we use the disentangled representation space from~\citep{naturalspeech3} that is pre-trained on 60,000 hours~\citep{librilight}. During the speech synthesis process, we adopt an encoder-decoder architecture~\citep{fastspeech2} as the backbone synthesis framework and integrate a high-quality non-autoregressive, confidence-based codec generator~\citep{maskgit,soundstorm,videomask} as the decoder.

We also identify and analyze the many-to-many issue in textual style-controllable TTS for the first time, that is, different textual style descriptions may correspond to the same audio, \textbf{while a single textual style description may be associated with varying degrees of a particular style for the same speaker}. For instance, the phrases ``The man speaks at a very rapid pace" and ``The man articulates his words with considerable speed" describe the same speech style, yet ``The man speaks at a very rapid pace" can also correspond to many audio clips exhibiting \textbf{different levels of high speaking rate}. To address this many-to-many issue in style control, we propose a novel module called \textbf{Style Mixture Semantic Density Sampling (SMSD)}. This module integrates the global semantic information of style control and utilizes sampling from a mixed distribution ~\citep{mdn1,mdn2} of style descriptions to achieve hierarchical control. Additionally, we incorporate a noise perturbation mechanism to further enhance style diversity. The design motivation and detailed architecture of SMSD are elaborated in Section~\ref{subsec:smsd}.

To comprehensively evaluate ControlSpeech's controllability, timbre similarity, audio quality, diversity, and generalization, we create a new open sourced dataset called \textbf{VccmDataset} based on TextrolSpeech~\citep{textrolspeech} to foster advancements in controllable TTS. In summary, our contributions are as follows:

\begin{itemize}[leftmargin=*,noitemsep]
    \item  We conduct detailed analysis of existing zero-shot TTS and style-controllable TTS models and identify their inability to simultaneously and independently control content, style, and timbre in a zero-shot setting. We propose the ControlSpeech to achieve independent control over these speech factors at the same time. 
    \item To the best of our knowledge, this is also the first work to identify and analyze the many-to-many issue in text style-controllable TTS, we propose a novel Style Mixture Semantic Density (SMSD) module. Furthermore, we investigate integrating various noise perturbation mechanisms within SMSD to enhance control diversity.
    \item We conduct comprehensive experiments and demonstrate that ControlSpeech exhibits comparable or state-of-the-art performance in terms of controllability, timbre similarity, audio quality, robustness, and generalizability. We also create a new open-source dataset VccmDataset tailored for style and timbre control at the same time.
\end{itemize}

\section{Related Work}
\label{sec:related_work}
In this section, we summarize previous studies on text prompt-based controllable TTS. Detailed discussions of discrete codec related to ControlSpeech are in Appendix~\ref{appendix related work}. 

\subsection{Text Prompt Based Controllable TTS}
Some recent studies propose to control speech style through natural text prompts. PromptTTS~\citep{prompttts} employs manually annotated text prompts to describe four to five attributes of speech (gender, pitch, speaking speed, volume, and emotion). 
InstructTTS~\citep{instructtts} employs a three-stage training approach to capture semantic information
from natural language style prompts as conditioning to the TTS system. Textrolspeech~\citep{textrolspeech} introduces an efficient architecture which treats textual controllable TTS as a language model task. PromptStyle~\citep{promptstyle} proposes a two-stage TTS approach for cross-speaker
style transfer with natural language descriptions based on VITS~\citep{vits}. PromptTTS 2~\citep{prompttts2} proposes an automatic description creation pipeline leveraging large language models (LLMs)~\citep{prompttts2sparks} and adopts a diffusion model to capture the one-to-many relationship. Audiobox~\citep{audiobox} propose a unified model based on flow-matching that
is capable of generating and controlling various audio modalities. While AudioBox supports multiple inputs, it does not decouple the speech prompt from the style prompt. Consequently, when there is a conflict between the styles in the speech prompt and the style text prompt, it significantly impacts the controllability. We also validate the necessity of decoupling in our ablation study presented in Table~\ref{table_codec}. It is noteworthy that existing style-controllable TTS models are either speaker-independent or can only control timbre using speaker IDs, without the capability for timbre cloning. The introduction of ControlSpeech expands the scope of the controllable TTS task. 

Furthermore, to the best of our knowledge, ControlSpeech is the first model to identify the many-to-many problem in the field of style control. It is worth noting that while PromptTTS 2~\citep{prompttts2} also identifies a one-to-many issue between style descriptions and audio, the one-to-many issue identified in PromptTTS 2 is fundamentally different from the one-to-many issue we identify in ControlSpeech. PromptTTS 2 attributes the one-to-many issue to the absence of the timbre information in the style descriptions, and thus employs a Q-former combined with a diffusion model to generate \textbf{the missing latent speech features}. In contrast, we argue that \textbf{the textual style descriptions themselves are inherently insufficient to capture the range of variations in one style}, leading to the one-to-many issue. 

\subsection{Zero-shot TTS}
Zero-shot speech synthesis refers to the ability to synthesize the voice of an unseen speaker based solely on a few seconds of audio prompt, also known as voice cloning. In recent months, with the advancement of generative large-scale models, a plethora of outstanding works have emerged. VALL-E~\citep{valle} leverages discrete codec representations and combines autoregressive and non-autoregressive models in a cascaded manner, preserving the powerful contextual capabilities of language models. NaturalSpeech 2~\citep{naturalspeech2} employs continuous vectors instead of discrete neural codec tokens and introduces in-context learning to a latent diffusion model. NaturalSpeech 3~\citep{naturalspeech3} proposes a TTS system with novel factorized diffusion models to generate natural speech in a zero-shot way, although Naturalspeech 3 also employs a disentangled codec representation, all its codec targets are generated with the same textual content. SpearTTS~\citep{speartts} and Make-a-Voice \citep{make-a-voive} utilize semantic tokens to reduce the gap between text and acoustic features. VoiceBox~\citep{le2023voicebox} is a non-autoregressive flow-matching model trained to infill speech, given audio context and text. Mega-TTS~\citep{megatts,megatts21,megatts22}, on the other hand, utilizes traditional mel-spectrograms, decoupling timbre and prosody and further modeling the prosody using an autoregressive approach. VoiceBox~\citep{le2023voicebox} and P-flow \citep{pflow} employ flowing models as generators, demonstrating robust generative performance. SoundStorm~\citep{soundstorm} and MobileSpeech~\citep{mobilespeech} utilize a non-autoregressive and mask-based iterative generation method, achieving an excellent balance between inference speed and generation quality. It is noteworthy that existing zero-shot TTS models (including NaturalSpeech3) are unable to achieve arbitrary language style \textbf{control and modify. ControlSpeech is the first TTS model capable of simultaneously and independently performing zero-shot timbre cloning and style control.}

\section{ControlSpeech}
\label{sec:method}

\begin{figure*}[t]
\centering
\includegraphics[height=6cm, width=16cm]{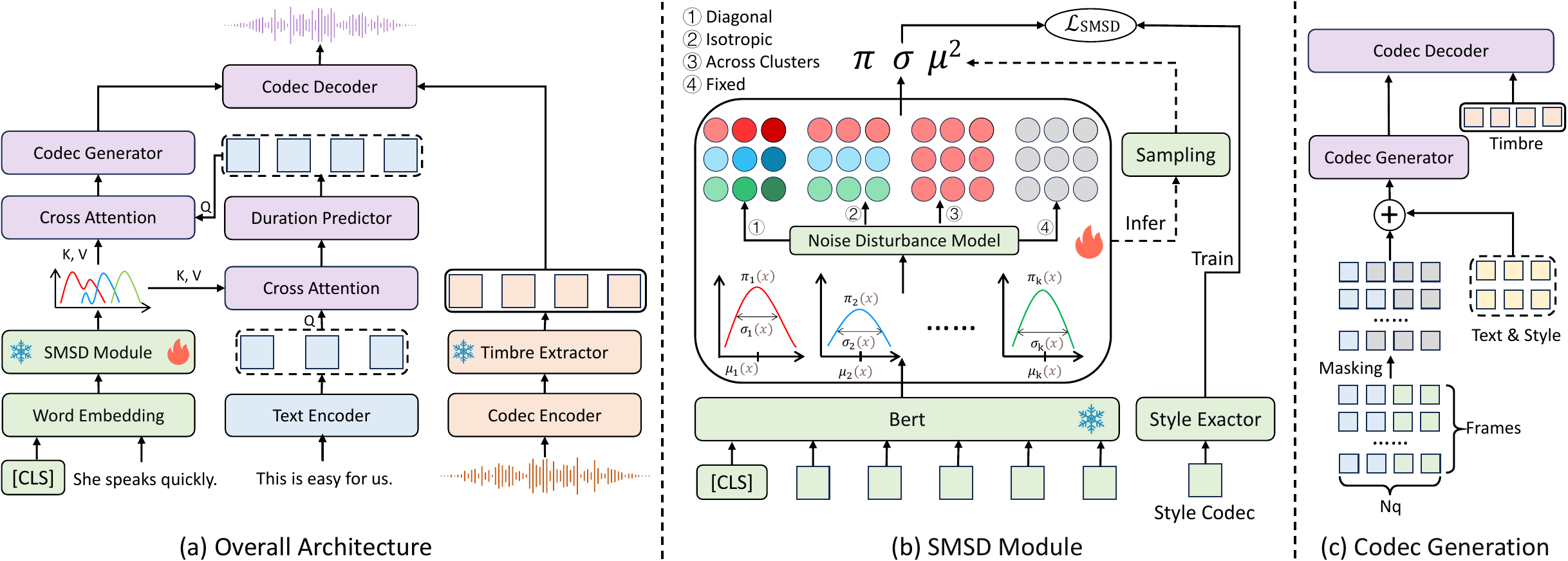}

\caption{Figure (a) depicts the overall architecture of ControlSpeech, which is an encoder-decoder parallel disentangled codec generation model. Figure (b) provides a detailed illustration of the SMSD module, which addresses the many-to-many problem in style control by sampling from the style mixture semantic distribution and incorporating an additional noise perturbator. Figure (c) shows the process of the codec generator. Through masking, the codec can generate discrete codec representations in a fully non-autoregressive manner.}
\label{arc2}
\end{figure*}

\subsection{Overall Architecture}
\label{subsec:overall_architecture}
 As illustrated in Figure~\ref{arc2} (a), ControlSpeech is fundamentally an encoder-decoder model \citep{mobilespeech} designed for parallel codec generation~\citep{soundstorm}. ControlSpeech employs three separate encoders to encode the input content prompt, style prompt, and speech prompt, respectively. Specifically, the content text is converted into phonemes and fed into the text encoder, while style text is prepended with the special [CLS] token and encoded at the word level using BERT's tokenizer~\citep{bert}. Meanwhile, the speech prompt is processed by the pre-trained codec encoder~\citep{naturalspeech3} and timbre extractor to capture the timbre information. In Figure~\ref{arc2}, the dashed box represents frame-level features, while the solid box represents global features. The Style Mixture Semantic Density (SMSD) module samples style text to generate the corresponding global style representations, which are then combined with text representations from the text encoder via a cross-attention module. The combined representations are then fed into the duration prediction model and subsequently into the codec generator, which is a non-autoregressive Conformer based on mask iteration and parallel generation. The timbre extractor is a Transformer encoder that converts the output of the speech encoder into a global vector, representing the timbre attributes. Given the input of a style description $X_{s}$, a content text $X_{c}$, and a speech prompt $X_{t}$, ControlSpeech aims to sequentially generate the corresponding style codec $Y_{s}$, content codec $Y_{c}$, and timbre embedding $Y_{t}$. These representations are then concatenated and upsampled into speech through the pre-trained codec decoder~\citep{naturalspeech3}.

\subsection{Codec Decoupling and Generation}
\label{subsec:decoupling}

\subsubsection{Decouple Content, Style, and Timbre}
ControlSpeech leverages the pre-trained disentangled representation space to separate different aspects of speech. 
We utilize FACodec~\citep{naturalspeech3} as our codec disentangler and timbre extractor module, since FACodec facilitates codec decoupling and is pre-trained on a large-scale, multi-speaker dataset, ensuring robust zero-shot TTS capabilities. Specifically, during the training process of ControlSpeech, we freeze the corresponding codec encoder to obtain downsampled compressed audio frames $h$ from the target speech $Y$. The frames $h$ are processed through the disentangling quantizer module and the timbre extractor module~\citep{naturalspeech3} to derive the original content codec $Y_{c}$, prosody codec $Y_{p}$, acoustic codec $Y_{a}$, and timbre information 
$Y_{t}$. Theoretically, after excluding the content $Y_{c}$ and timbre information $Y_{t}$, the remaining representation collectively is treated as the style codec $Y_{s}$. In practice, we concatenate the prosody codec $Y_{p}$ and the acoustic codec $Y_{a}$ along the channel dimension to obtain the corresponding style codec $Y_{s}$, as follows:
\begin{equation}
    Y_{s} = concat(Y_{p},Y_{a})
\end{equation}

\subsubsection{Codec Generation Process}
\label{subsec:generation_process}
The codec generation comprises two stages.

\textbf{In the first stage}, based on the paired text and speech data $\left \{ X,Y_{codec} \right \} $, where $X=\left \{ x_{1},x_{2},x_{3},\cdots, x_{T}\right \} $ represents the cross-attention fusion of the global style representations and the aligned text representations, and $Y_{codec}$ denotes the speech representations through vector quantization, formulated as follows:
\begin{equation}
    Y_{codec} = concat(Y_{s},Y_{c}) = C_{1:T,1:N}\in  \mathbb{R}^{T\times N}
\end{equation}
\noindent where $T$ denotes the downsampled utterance length, which is equal to the text length extended by the duration predictor. $N$ represents the number of channels for every frame. The row vector of each acoustic code matrix $C_{t,1:N}$ represents the $N$ codes for frame $t$, and the column vector of each acoustic code matrix $C_{1:T,i}$ represents the $i$-th codebook sequence (the length is $T$), where $i\in \left \{ 1,2,\cdots,N \right \} $.
 Following VALL-E \citep{valle}, in the training process of ControlSpeech, we randomly select the $i$-th channel $C_{1:T,i}$ for training. For the generation of the $i$-th channel $P(C_{1:T,i}\mid X_{1:T};\theta)$, as illustrated in Figure~\ref{arc2} (c), we employ a mask-based generative model as our parallel decoder. We sample the mask $M_{i}\in \left \{ 0,1 \right \} ^T$ according to a cosine schedule~\citep{maskgit} for codec level $i$, specifically, sampling the masking ratio $p=\cos(u^{'})$ where $u^{'}\sim \mathcal{U}\left [  0,\frac{\pi }{2}  \right ] $. and the mask $M_{i} \sim  Bernoulli(p) $. Here, $M_{i}$ represents the portion to be masked in the $i$-th level, while $\bar{M_{i}} $ denotes the unmasked portion in the $i$-th level. As shown in Figure~\ref{arc2} (c), the prediction of this portion $C_{1:T,i}$ is refined based on the prompt $j (j<i)$ channels $C_{1:T,<i}$, and the concatenation of the target text $X_{1:T}$ and the unmasked portion of the $i$-th channel $\bar{M_{i}}C_{1:T,i}$. Therefore, the prediction for this part can be specified as $
     P(C_{1:T,i}\mid X_{1:T};\theta) = P(M_{i}C_{1:T,i}\mid C_{1:T,<i},X_{1:T},\bar{M_{i}}C_{1:T,i};\theta)$


 \textbf{In the second stage}, as illustrated in Figure~\ref{arc2} (c), following AdaSpeech \citep{adaspeech}, we utilize a conditional normalization layer to fuse the previously obtained $Y_{codec}$ and the global timbre embedding $Y_{t}$, resulting in $Y^{'}$. This result $Y^{'}$ is then processed by the pre-trained codec decoder~\citep{naturalspeech3} to generate the final speech output $Y$. Specifically, we first use two simple linear layers $W_{\gamma }$ and $W_{\beta }$, which take the global timbre embedding $Y_{t}$ as input and output the scale vectors $\gamma$ and bias vectors $\beta$ respectively. These lightweight, learnable scale vectors $\gamma$ and bias vectors $\beta$ are then fused with $Y_{codec}$. This process can be represented by the following formula:
 \begin{equation}
     Y = CodecDecoder(W_{\gamma}Y_{t}\frac{Y_{codec}-\mu_{c} }{{\sigma_{c} }^{2}  } +W_{\beta }Y_{t})
 \end{equation}
\noindent where $\mu_{c}$ and ${\sigma_{c}} ^{2} $ are the mean and variance of the hidden representation of $Y_{codec}$.

\subsection{The Style Mixture Semantic Density (SMSD) Module}
\label{subsec:smsd}
We identify a \textbf{many-to-many} relationship between style text descriptions and their corresponding audio. Specifically, different style descriptions can correspond to the same audio sample (that is, \textbf{many-to-one}), while a single style description may correspond to multiple audio samples with varying degrees of the same style (that is, \textbf{one-to-many}). More precisely, the many-to-one relationship arises because multiple textual descriptions can refer to the same style of speech. For example, both ``Her speaking speed is considerably fast'' and ``Her speech rate is remarkably fast'' can refer to the ``fast-speed'' speech style and could correspond to the same audio sample. On the other hand, the one-to-many relationship occurs because a single textual description is unable to capture the \textbf{varying degrees} of a style. For instance, if we divide the tempo of different speech into 100 levels, any speech with the tempo above 70 may be considered as ``fast-speed''. As a result, the text description suggesting ``fast speed'' could correspond to different audio samples with speech rates of 75, 80, or even 90 for the same speaker.

To address the many-to-many issue in style control, we propose the Style Mixture Semantic Density (SMSD) module. To address the many-to-one issue, similar to previous approaches~\citep{prompttts,promptstyle}, we utilize a pre-trained BERT model within the SMSD module to extract the semantic representation ${X_{s}}^{'}$ from style descriptions, thereby aligning different style texts into the same semantic space and enhancing generalization of out-of-domain style descriptions. To address the one-to-many issue, we observe that addressing this phenomenon of a single style description corresponding to multiple audio with varying degrees of style closely aligns with the motivation of mixture density networks (MDN). We hypothesize that ${X_{s}}^{'}$ as the semantic representation of style can be considered as a global mixture of Gaussian distributions, where different Gaussian distributions represent varying degrees of a particular style. During training, each independent Gaussian distribution is multiplied by a corresponding learnable weight and then summed. By constraining the $KL$ divergence between the style representation distribution of the target audio and the summed mixture density distribution, we establish a one-to-one correspondence between the style text and the target audio. This approach also enhances the diversity of style control directly with the text descriptions. During inference, we sample from the mixture of style semantic distributions to obtain an independent Gaussian distribution, with each sampled distribution reflecting different degrees of the same style. Additionally, to further enhance the diversity of style control, we incorporate a noise perturbation module within the MDN network of SMSD in ControlSpeech. The noise perturbation module controls the isotropy of perturbations across different dimensions.

Specifically, one raw style prompt $X_{s} = [X_{1},X_{2},X_{3},\cdots,X_{L}]$ is prepended with a $\left[ CLS \right]$ token, then converted into word embedding, and fed into the BERT model, where $L$ denotes the length of the style prompt. The hidden vector corresponding to the $\left[ CLS \right]$ token is regarded as the global style semantic representation ${X_{s}}^{'}$, which guides generation and sampling of subsequent modules. Based on the MDN network~\citep{mdn1,mdn_github,dumdn3}, we aim to regress the target style representation ${Y_{s}}^{'} \in \mathbb R^{d}$, using the style semantic input representation ${X_{s}}^{'} \in \mathbb R^{n}$ as covariates, where $d$ and $n$ are the respective dimensions. We model the conditional distribution as a mixture of Gaussian distribution, as follows:
\begin{equation}
    P_{\theta }({Y_{s}}^{'}|{X_{s}}^{'}) = \sum_{k=1}^{K}\pi_{k}\mathcal{N}(\mu^{(k)},{\sigma^{2}}^{(k)} ) 
\end{equation}
where $K$ is a hyperparameter as the number of independent Gaussian distribution, and other mixture distribution parameters $\pi_{k}$, $\mu^{k}$, ${\sigma^{2}}^{(k)} $ are output of a neural MDN network $f_{\theta}$ based on the input style semantic representation ${X_{s}}^{'}$, as follows:
\begin{equation}
    \pi \in \Delta ^{K-1}, \mu^{(k)} \in \mathbb{R}^{d},{\sigma ^{2}}^{(k)} \in S_{+}^{d} = f_{\theta }({X_{s}}^{'})
\end{equation}
Note that the sum of the mixture weights is constrained to 1 during the training phase, which is achieved by applying a softmax function on the corresponding neural network output $\alpha_{k}$, as follows:
\begin{equation}
    \pi_{k} = \frac{exp(a_{k})}{\sum_{k=1}^{K} exp(a_{k})} 
\end{equation}
To further enhance the diversity of style control, we design a specialized noise perturbation module within the SMSD module to constrain the noise model. As illustrated by the circles within the SMSD module in Figure~\ref{arc2} (b), this noise perturbation module regulates the isotropy of perturbations $\varepsilon$ across different dimensions in variance ${\sigma^{2}}^{(k)}$. The four types of perturbations from left to right in Figure \ref{arc2} (b) are as follows:

\begin{itemize}[leftmargin=*,noitemsep]
    \item \textbf{Fully factored}: ${\sigma^{2}}^{(k)} = f_{\theta }({X_{s}}^{'})+f_{\theta}(\varepsilon ) = diag({\sigma^{2}}^{(k)})\in \mathbb R_{+}^{d}$, which predicts the noise level for each dimension separately.
    \item \textbf{Isotropic}: ${\sigma^{2}}^{(k)} = f_{\theta }({X_{s}}^{'})+f_{\theta}(\varepsilon ) = {\sigma^{2}}^{(k)}I \in \mathbb R_{+}$, which assumes the same noise level for each dimension over $d$.
    \item \textbf{Isotropic across clusters}: ${\sigma^{2}}^{(k)} = f_{\theta }({X_{s}}^{'})+f_{\theta}(\varepsilon ) = \sigma^{2}I \in \mathbb R_{+}$, which assumes the same noise level for each dimension over $d$ and cluster.
    \item \textbf{Fixed isotropic} is the same as Isotropic across clusters but does not learn $\sigma ^{2}$.
\end{itemize}

As shown in the experimental results in Appendix~\ref{appendix noise}, \textit{isotropic across clusters} outperforms the other types for striking a balance between accuracy and diversity and is used as the mode for noise perturbation. We obtain more robust mean, variance, and weight parameters for the mixture of Gaussian distributions with the noise perturbation module. The training objective of the SMSD module is the negative log-likelihood of the observation ${Y_{s}}^{'}$ given its input ${X_{s}}^{'}$. The loss function is formulated as $\mathcal L_{SMSD}=-logsumexp_{k}(log\pi_{k}-\frac{1}{2}\left \| \frac{{Y_{s}}^{'}-\mu^{(k)}}{\sigma }  \right \|^{2} )$.
Details for deriving the non-convex $\mathcal L_{SMSD}$ are in Appendix~\ref{appendix smsdloss}.



\subsection{Training and Inference}
\label{subsec:training_inference}
During the training process, the duration predictor is optimized using the mean square error loss, with the extracted duration serving as the training target. We employ the Montreal Forced Alignment (MFA) tool~\citep{mfa} to extract phoneme durations, and denote the loss for the duration predictor as $\mathcal L_{dur}$. The codec generator module is optimized using cross-entropy loss. We randomly select a channel for optimization and denote this loss as $\mathcal L_{codec}$. In the SMSD module, the target style representation ${Y_{s}}^{'}$ is the global style representation obtained by passing style codec $Y_{s}$
  through the style extractor. During training, we feed the ground truth style representation ${Y_{s}}^{'}$ and the ground truth duration into the codec generator and duration predictor, respectively. The overall loss $\mathcal L$ for ControlSpeech is the sum of losses:

  \begin{equation}
      \mathcal L = \mathcal L_{codec} + \mathcal L_{dur} + \mathcal L_{SMSD}
  \end{equation}

During the inference stage, we initiate the process by inputting the original stylistic descriptor $X_{s}$ into the BERT module to obtain the style semantic representation ${X_{s}}^{'}$, and then input ${X_{s}}^{'}$ into the SMSD module to obtain the corresponding $\pi$, $\mu$ and $\sigma^{2}$. By directly sampling ${X_{s}}^{'}$, we can derive the predicted style distribution. Subsequently, we iteratively generate discrete acoustic tokens by incorporating the predicted style into the text state and employing the confidence based sampling scheme~\citep{maskgit,soundstorm}. Specifically, we perform multiple forward passes, and at each iteration $j$, we sample candidates for the masked positions. We then retain $P_{j}$ candidates based on their confidence scores, where $P_{j}$ follows a cosine schedule. Finally, by integrating the timbre prompt through the condition normalization layer and feeding it into the codec decoder, we generate the final speech output.

\section{Experiments}
\label{sec:experiments}

\subsection{Experimental Setup}
\label{subsec:experimental_setting}

\paragraph{VccmDataset.}
To the best of our knowledge, there is no large-scale TTS dataset that includes both text style prompts and speaker prompts. We build upon the TextrolSpeech dataset~\citep{textrolspeech} and create \textbf{VccmDataset}. Based on TextrolSpeech, we optimize the pitch distribution, label boundaries, the dataset splits, and then select new test sets.  
Specifically, we use LibriTTS and the emotional data from TextrolSpeech as the base databases, and annotate each speech sample with five attribute labels: gender, volume, speed, pitch, and emotion. We use the gender labels available in the online metadata. Regarding volume, we compute the L2-norm of the amplitude of each short-time Fourier transform frame. We utilize the Montreal forced alignment tool~\citep{mfa} to extract phoneme durations and silence segments. Subsequently, we calculate the average duration of each phoneme within voiced segments for the speaking speed.  The Parselmouth 3 tool \footnote{\url{https://github.com/YannickJadoul/Parselmouth}} is employed to extract fundamental frequency (f0) and calculate the geometric mean across all voiced regions as pitch values. We partition speech samples into 3 categories (high/normal/low) according to the proportion of speed, pitch, and volume values respectively. Considering the close proximity of attribute values of speech samples between adjacent categories, we exclude the 5$\%$ of data samples at the boundaries of each interval for each attribute. Particularly, We use gender-specific thresholds to bin the pitch into three different levels. After obtaining more accurate labels through these procedures, we align each audio segment with the corresponding style description text in TextrolSpeech based on the labeled attributes to obtain the VccmDataset.
We then select four distinct test sets from VccmDataset, namely, \textbf{\textit{test set A}}, \textbf{\textit{test set B}}, \textbf{\textit{test set C}}, \textbf{\textit{test set D}}. Details of the VccmDataset test sets are in Appendix~\ref{appendix vccmdataset test set}.

\paragraph{Baselines.}
To ensure a fair comparison of the actual performance of various models, we reimplement several SOTA style-controllable models, including PromptStyle \citep{promptstyle}, Salle \citep{textrolspeech}, InstructTTS \citep{instructtts}, and PromptTTS 2 \citep{prompttts2}, to serve as primary comparative models for evaluating the controllability of ControlSpeech. For the comparison of voice cloning effectiveness, we reimplement the VALL-E model~\citep{valle} and the MobileSpeech model~\citep{mobilespeech}, which are representatives of the autoregressive paradigm and the parallel generation paradigm, respectively. All reproduced baseline will be also made publicly. 

\paragraph{Evaluation Metrics and Experimental Settings.}
For objective evaluations, we adopt the common metrics used in prior works~\citep{prompttts,textrolspeech,prompttts2}. To evaluate the model's style controllability, we use accuracy of pitch, speaking speed, volume, emotion as the metrics, which measures the correspondence between the style factors in the output speech and those in the prompts. We evaluate timbre similarity (Spk-sv) between the original prompt and the synthesized speech, and evaluate speech synthesis accuracy and robustness by using an ASR system to transcribe the synthesized speech and computing word error rate (WER) against the content prompt. For subjective evaluations, we conduct mean opinion score (MOS) evaluations on the test set to measure audio naturalness via crowdsourcing. We further analyze MOS in two aspects: MOS-Q (Quality, assessing clarity and naturalness of the duration and pitch) and MOS-S (Speaker similarity). We also design new subjective MOS metrics: \textbf{MOS-TS} (Timbre similarity), \textbf{MOS-SD} (Style diversity), and \textbf{MOS-SA} (Style accuracy). Details of the evaluation metrics, experimental settings, and specifics of model architecture are provided in Appendix \ref{appendix metrics}, \ref{appendix training}, and \ref{appendix modelarc}, respectively.

\begin{table*}[t]
\caption{The \textbf{style controllability} evaluation results  of style-controlled models on VccmDataset \textbf{\textit{test set A}}. \textit{Pitch}, \textit{Speed}, \textit{Volume}, \textit{Emotion} denote accuracy of the style. $\pm$ denotes standard deviation.}

\centering
\begin{adjustbox}{width=\textwidth}
\begin{tabular}{c|ccccccccc}
\hline
Model & Clone Timbre & Control Style  & Pitch $\uparrow$ & Speed $\uparrow$  & Volume $\uparrow$ & Emotion $\uparrow$ & WER $\downarrow$ & Spk-sv $\uparrow$ & MOS-Q $\uparrow$ \\
\hline
GT Codec & - & - & 0.954 & 0.885 & 0.977 & 0.758 & 2.6 & 0.96 & 4.25$\pm$0.10\\ 
Salle & $\times $ & $\surd $ & 0.788 & 0.756 & 0.831 & 0.389 & 5.5 & - & 3.52$\pm$0.14\\ 
PromptStyle& $\times $ & $\surd $ & 0.831 & 0.786& 0.787& 0.366 & 3.3 & 0.84 & 3.74$\pm$0.11\\ 
InstructTTS& $\times $ & $\surd $ & 0.849& 0.761 & 0.822 & 0.412  & 3.0 & 0.86 &3.81$\pm$0.12\\ 
PromptTTS 2& $\times $ & $\surd $ & \textbf{0.867} & 0.785 & 0.825 & 0.406& 3.1 & - & 3.83$\pm$0.11\\ 
ControlSpeech (\textbf{Ours}) & $\surd $ & $\surd $ & 0.833 & \textbf{0.829} & \textbf{0.894} & \textbf{0.557} & \textbf{2.9} & \textbf{0.89} & \textbf{3.91$\pm$0.09}\\ 
\hline
\end{tabular}
\end{adjustbox}
\label{table1}
\end{table*}

\subsection{Results and Discussions}
\label{subsec:evaluations}



\paragraph{Evaluation on style controllability.}
We first compare the performance of ControlSpeech with various SOTA models on the style controllability task. The evaluation is conducted on the 1,500-sample VccmDataset \textbf{\textit{test set A}}. To eliminate the influence of timbre variations on the controllability results of ControlSpeech, we use the ground truth (GT) timbre as the prompt for ControlSpeech. We compare the controllability of the models using pitch accuracy, speed accuracy, volume accuracy, and emotion accuracy. Additionally, we measure the audio quality generated by the models using WER, timbre similarity (Spk-sv), and MOS-Q. Results are shown in Table~\ref{table1}, and we drew the following conclusions: \textbf{1)} Comparing ControlSpeech with other baselines on controllability metrics, we find that, except for pitch accuracy, ControlSpeech achieves best results in volume, speed, and emotion classification accuracy. Upon analyzing the synthesized audio of ControlSpeech, we attribute the degraded pitch accuracy to the difficulty arising from simultaneously controlling different timbres and styles. \textbf{2)} In terms of Spk-sv, MOS-Q, and WER metrics, the audio generated by ControlSpeech demonstrates best timbre similarity, audio quality, and robustness.


\paragraph{Evaluation on the timbre cloning task.}

\begin{table}[htbp]
\caption{The \textbf{timbre cloning} results of different zero-shot models on the VccmDataset \textbf{\textit{test set B}}.}
\centering
\begin{adjustbox}{width=0.48\textwidth}
\begin{tabular}{c|ccccc}
\hline
Model & Clone Timbre & Control Style  & WER $\downarrow$ & MOS-Q $\uparrow$  & MOS-S $\uparrow$  \\
\hline
GT Codec & - & - & 2.3  & 4.21$\pm$0.14 & 4.29$\pm$0.12\\ 
VALL-E & $\surd $ & $\times $ & 6.7 & 3.76$\pm$0.13 & 3.89$\pm$0.13\\ 
MobileSpeech & $\surd $ & $\times $ & 4.1 & 3.94$\pm$0.09& \textbf{4.01$\pm$0.11} \\ 
ControlSpeech (\textbf{Ours}) & $\surd $ & $\surd$ & \textbf{3.3} & \textbf{3.95$\pm$0.12} & 3.96$\pm$0.14 \\ 
\hline
\end{tabular}
\end{adjustbox}
\label{table2}
\end{table}
To evaluate the timbre cloning capability of ControlSpeech in an out-of-domain speaker scenario, we compare the performance of ControlSpeech with SOTA models such as VALL-E and MobileSpeech on the out-of-domain speaker test set (\textbf{\textit{test set B}}). The experimental results are shown in Table~\ref{table2}. We observe that in terms of the robustness metric (WER), the zero-shot TTS systems that are trained on small datasets perform worse than ControlSpeech. We attribute these performance gains of ControlSpeech to its pre-trained speaker prompt component. Additionally, in terms of the MOS-Q and MOS-S metrics, we find that ControlSpeech also maintains performance comparable to zero-shot TTS systems on the timbre cloning task.

\paragraph{Evaluation on addressing the many-to-many issue.}

To better evaluate the performance of style-controllable models on addressing the many-to-many issue, we compare ControlSpeech with controllable baseline models on the VccmDataset \textbf{\textit{test set D}}. Results are shown in Table~\ref{table4}. We find that ControlSpeech markedly outperforms PromptStyle and InstructTTS on both MOS-SA (style accuracy) and MOS-SD (style diversity) metrics. This suggests that the unique SMSD module in ControlSpeech enables the model to synthesize both accurate and diverse speech.

\begin{table}[htbp]
\caption{The results under \textbf{many-to-many style control conditions} on VccmDataset \textbf{\textit{test set D}}. MOS-TS, MOS-SA, MOS-SD measure timbre stability, accuracy and diversity of style generation.}
\centering
\begin{adjustbox}{width=0.48\textwidth}
\begin{tabular}{c|ccc}
\hline
Model  & MOS-TS $\uparrow$ & MOS-SA $\uparrow$  & MOS-SD$\uparrow$   \\
\hline
PromptStyle & 3.81$\pm$0.10 & 3.45$\pm$0.13 & 3.53$\pm$0.12\\ 
InstructTTS & 3.89$\pm$0.12 & 3.57$\pm$0.11 & 3.48$\pm$0.14 \\ 
ControlSpeech w/o SMSD & 3.95$\pm$0.08 & 3.59$\pm$0.09  & 3.66$\pm$0.11\\
ControlSpeech & \textbf{4.01$\pm$0.10} & \textbf{3.84$\pm$0.12} &\textbf{4.05$\pm$0.09}  \\ 
\hline
\end{tabular}
\end{adjustbox}
\label{table4}
\end{table}

\subsection{Ablation Studies}
\label{subsec:ablation}
We validate the necessity of the codec decoupl ingand the SMSD module. We also investigate the impact of hyperparameters for mixed distributions and various noise models in Appendix \ref{appendix mix} and \ref{appendix noise}.

\begin{table}[htbp]
\caption{An ablation experiment on impact of \textbf{codec decoupling} on the VccmDataset \textbf{\textit{test set A}}.}
\centering
\begin{adjustbox}{width=0.48\textwidth}
\begin{tabular}{c|cccc}
\hline
Model  & Pitch $\uparrow$ & Speed $\uparrow$  & Volume $\uparrow$ & Emotion $\uparrow$ \\
\hline
ControlSpeech w/o decoupling  & 0.492 & 0.517 & 0.582 & 0.237 \\
ControlSpeech & \textbf{0.833} & \textbf{0.829} & \textbf{0.894} & \textbf{0.557}\\ 
\hline
\end{tabular}
\end{adjustbox}
\label{table_codec}
\end{table}

\textbf{Decouple codec.} To analyze the impact of decoupling, we maintain the main framework of ControlSpeech and directly encode the speech prompt and style prompt using the frozen speech encodec encoder and style encoder (replicated from the structure of the text encoder) respectively, then feed them into the codec generator through cross attention. We denote this model as \textbf{ControlSpeech w/o decoupling}. As shown in Table~\ref{table_codec}, ControlSpeech w/o decoupling performs substantially worse in controllability compared to ControlSpeech, suggesting that the speech prompt and style prompt indeed may interfere with each other.

\textbf{The SMSD module.}  We replace the SMSD module with a style encoder (replicated from the structure of the text encoder) and denote this model as \textbf{ControlSpeech w/o SMSD}. As shown in Table~\ref{table4}, ControlSpeech w/o SMSD performs markedly worse in terms of MOS-SA and MOS-SD compared to ControlSpeech, which strongly validates that the SMSD module enables more fine-grained control of the model's style and increases style diversity through style sampling. We also visualize the distribution of the SMSD under varying pitch/speed/volume (details in Appendix~\ref{appendix show smsd}). 

\section{Conclusion}
\label{sec:conclusion}
In this paper, we present ControlSpeech, the first TTS system capable of simultaneously performing zero-shot timbre cloning and zero-shot style control independently. Additionally, we identify a many-to-many problem in style control and design a unique SMSD module. We will also open source VccmDataset to foster community development.

\section{Acknowledgments}
 This work was supported in part by the National Natural Science Foundation of China under Grant No.62222211 and No.U24A20326

\section*{Limitations}
\label{appendix future}
In this work, we introduce ControlSpeech, the first TTS system capable of simultaneously cloning timbre and controlling style independently. While ControlSpeech has demonstrated competitive controllability and cloning capabilities, there remains considerable scope for further research and improvement based on the current framework.

    \paragraph{Larger Training Datasets.} The field of style-controllable TTS demands larger training datasets. Although TextrolSpeech and our VccmDataset have established a foundation, we hypothesize that achieving more advanced speech controllability may require datasets comprising tens of thousands of hours of speech with style descriptions.
    
    \paragraph{Exploring Generative Models.} In this work, we experiment with decoupled codecs and non-autoregressive parallel generative models. In future research, we plan to explore a broader range of generative model architectures and audio representations.

\bibliography{custom}

\appendix


\section{Related work}
\label{appendix related work}

\subsection{Acoustic Codec Models}
In recent times, neural acoustic codecs \citep{soundstream,encodec,dac} have demonstrated remarkable capabilities in reconstructing high-quality
audio at low bitrates. Typically, these methods employ
an encoder to extract deep features in a latent space, which are subsequently quantized before being fed into the decoder. To elaborate, Soundstream~\citep{soundstream} utilizes a model architecture comprising a fully convolutional encoder/decoder network and a residual vector quantizer (RVQ) to effectively compress speech. Encodec~\citep{encodec} employs a streaming encoder-decoder architecture with a quantized latent space, trained in an end-to-end fashion. AudioDec~\citep{audiodec} has demonstrated the importance of discriminators. PromptCodec~\citep{promptcodec} enhances representation capabilities through additional input prompts. DAC~\citep{dac} significantly improves reconstruction quality through techniques like quantizer dropout and a multi-scale STFT-based discriminator. Vocos~\citep{vocos} eliminates codec noise artifacts using a pre-trained Encodec with an inverse Fourier transform vocoder. HILCodec~\citep{hilcodec} introduces the MFBD discriminator to guide codec modeling. APCodec~\citep{hilcodec} further enhances reconstruction quality by incorporating ConvNextV2 modules in the encoder and decoder. HiFi-Codec~\citep{hificodec} proposes a parallel GRVQ structure, achieving good speech reconstruction with just four quantizers. Language-Codec~\citep{languagecodec} introduces the MCRVQ mechanism to evenly distribute information across the first quantizer, also requiring only four quantizers for excellent performance across various generative models. Single-Codec~\citep{singlecodec} designs additional BLSTM, hybrid sampling, and resampling modules to ensure basic performance with a single quantizer, though reconstruction quality still needs improvement. TiCodec~\citep{ticodec} models codec space by distinguishing between time-independent and time-dependent information. FACodec~\citep{naturalspeech3} further decouples codec space into content, style, and acoustic detail modules.
Additionally, recognizing the importance of semantic information in generative models, recent efforts have begun integrating semantic information into codec models. RepCodec~\citep{repcodec} learns a vector quantization codebook by reconstructing speech representations from speech encoders like HuBERT. SpeechTokenizer~\citep{speechtokenizer} enriches the semantic content of the first quantizer through semantic distillation. FunCodec~\citep{funcodec} makes semantic tokens optional and explores different combinations. SemanticCodec~\citep{semanticodec} is based on quantized semantic tokens and further reconstructs acoustic information using an audio encoder and diffusion model. WavTokenizer~\citep{ji2024wavtokenizer} represents the latest state-of-the-art codec model, capable of reconstructing high-quality audio using only forty discrete codebooks. \textbf{Given that ControlSpeech requires disentangled discrete audio representations that are pre-trained on large-scale multi-speaker data, we select FACodec~\citep{naturalspeech3} as the tokenizer for ControlSpeech}.

\section{Distribution visualization}
\label{appendix show smsd}

\begin{figure}[htbp]
\centering
\includegraphics[height=2.5cm, width=8cm]{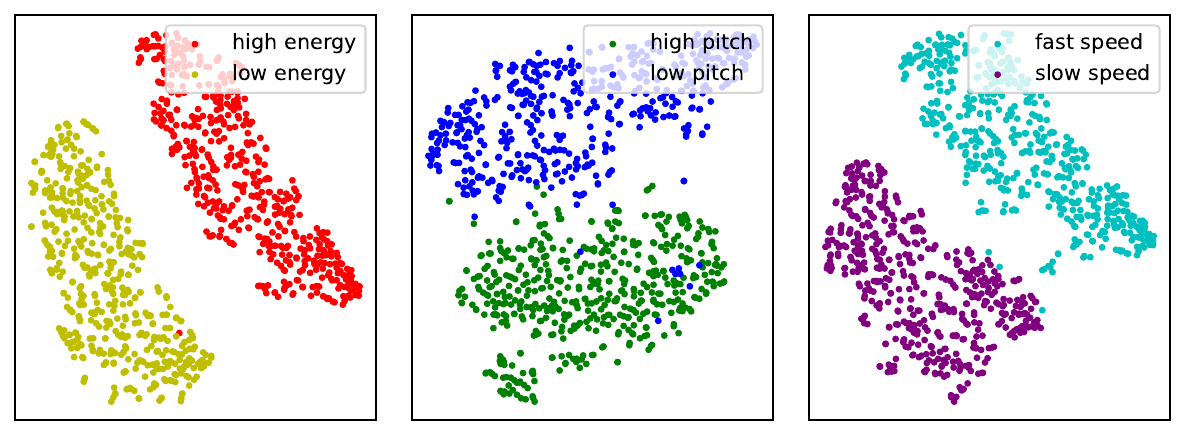}
\caption{The t-SNE visualization of  mixture density distribution after the SMSD module.}
\label{arc_smsd}
\end{figure}

In this section, we visualize the distribution of the SMSD mixed density network. As shown in Figure~\ref{arc_smsd}, we select the original style descriptions from TextrolSpeech and visualize the distributions produced by the SMSD module under three experimental settings: varying pitch (high/low), speech rate (fast/slow), and volume (high/low). Each experimental setting includes 1,000 different style descriptions, with other factors held constant. For example, in the speech rate experiment, both pitch and volume descriptions are set to ``normal." We employ t-SNE for dimensionality reduction of the features. Our results show that the SMSD module effectively distinguishes between different types of styles, and the mixed density distribution is not confined to a small region, indicating that the style control module exhibits substantial diversity.

\section{VccmDataset test set}
\label{appendix vccmdataset test set}
To further validate ControlSpeech's ability to simultaneously control style and clone speaker timbre, we create four types of test sets in the VccmDataset: the main test set (\textbf{\textit{test set A}}), the out-of-domain speaker test set (\textbf{\textit{test set B}}), the out-of-domain style test set (\textbf{\textit{test set C}}), and the special case test set (\textbf{\textit{test set D}}). Each test set corresponds to four experiments: style controllability experiments, out-of-domain speaker cloning experiments, out-of-domain style controllability experiments, and many-to-many style control experiments, respectively. We randomly select 1,500 audio samples as the ControlSpeech main test set (\textbf{\textit{test set A}}) and match the corresponding prompt voice based on speaker IDs. Additionally, to evaluate ControlSpeech's performance on out-of-domain timbre and styles, we further filter an appropriate test set (speakers that are not present in the training set) and enlist language experts to compose style descriptions distinct from those in TextrolSpeech. Using these two methods, we generate the out-of-domain speaker test set (\textbf{\textit{test set B}}) and the out-of-domain style test set (\textbf{\textit{test set C}}). The \textbf{\textit{test set B}} consists of 1,086 test utterances, and we ensure that none of the speakers in test set B appear in the training set. The special case test set (\textbf{\textit{test set D}}) is designed to evaluate the model's performance under many-to-many style control conditions. Firstly, we select four groups of speakers, each of whom is matched with 60 different style descriptions while the content text remains fixed. This particular set of test samples is referred to as \textbf{\textit{test set D1}}. We further select six distinct style descriptions paired with 50 different timbre prompts, with pitch, speed, and volume labels set to the following combinations: normal, fast, normal; normal, slow, normal; high, normal, normal; low, normal, normal; normal, normal, high; and normal, normal, low, respectively. This set of special test samples is referred to as \textbf{\textit{test set D2}}.

\section{Evaluation metrics}
\label{appendix metrics}
For objective evaluations, we adopt the metrics used in prior works~\citep{prompttts,textrolspeech,prompttts2}. To evaluate the model's style controllability, we use accuracy as the metric, which measures the correspondence between the style factors in the output speech and those in the prompts. The accuracy of pitch, speaking speed, and volume is calculated using signal processing tools. We fine-tune the official version of the Emotion2vec model~\citep{emotion2vec} on the emotional dataset of VccmDataset, and compute the speech emotion classification accuracy with the fine-tuned model. To evaluate timbre similarity (Spk-sv) between the original prompt and the synthesized speech, we utilize the base-plus-sv version of WavLM \citep{wavlm}. For Word Error Rate (WER), we use an ASR model \footnote{\url{https://huggingface.co/facebook/hubert-large-ls960-ft}} to transcribe the synthesized speech. This ASR model is a CTC-based HuBERT pre-trained on
Librilight and fine-tuned on the 960 hours training set of LibriSpeech. For subjective evaluations, we conduct mean opinion score (MOS) evaluations on the test set to measure audio naturalness via crowdsourcing. We randomly select 30 samples from the test set of each dataset for subjective evaluation, and each audio sample is listened by at least 10 testers. We analyze the MOS in two aspects: MOS-Q (Quality, assessing clarity and naturalness of the duration and pitch) and MOS-S (Speaker similarity).

Furthermore, for the evaluation of style-controllable many-to-many scenarios in the \textbf{\textit{test set D}}, we design new subjective MOS metrics: MOS-TS (Timbre Similarity), MOS-SD (Style Diversity), and MOS-SA (Style Accuracy). Specifically, the MOS-TS metric is used to assess whether the timbre remains stable across 60 different style descriptions for four speakers on the \textbf{\textit{test set D1}}. The MOS-SA and MOS-SD metrics represent the accuracy and diversity of style control for each style description respectively on the \textbf{\textit{test set D2}}.

\section{Training and Inference Settings}
\label{appendix training}
ControlSpeech is trained on VccmDataset using 8 NVIDIA A100 40G GPUs with each batch accommodating 3500 frames of the discrete codec. We optimize the models using the AdamW optimizer with parameters $\beta _{1}$ = 0.9 and $\beta_{2}$ = 0.95. The learning rate is warmed up for the first 5k updates, reaching a peak of $5\times 10^{-4}$, and then linearly decayed. We utilize the open-source FACodec's voice conversion version as the codec encoder and decoder for ControlSpeech. The style-controllable baseline models are trained on the same VccmDataset training set to eliminate potential biases. We utilize a pre-trained BERT~\citep{bert} model consisting of 12 hidden layers with 110M parameters. For the implementation of the basic MDN network model, we largely follow the approach described in~\citep{mdn_github}.

\section{Model Architecture in ControlSpeech}
\label{appendix modelarc}
Following~\citep{naturalspeech3}, the basic architecture of codec encoder and codec decoder follows~\citep{dac} and employs
the SnakeBeta activation function~\citep{bigvgan}. The timbre extractor consists of several conformer~\citep{conformer} blocks. We use $N_{{q}_{c}}=2$, $N_{{q}_{p}}=1$, $N_{{q}_{d}}=3$ as the number of quantizers for each of the three FVQ $Q^{c}$, $Q^{p}$, $Q^{d}$, the codebook size for all the quantizers is 1024. Text encoder and variance adaptor share the similar architecture which comprises several FFT blocks or attention layers as used by FastSpeech2~\citep{fastspeech2}. The Style Extractor is a module comprising both convolutional and LSTM networks from FACodec~\citep{naturalspeech3} and outputs a 512-dimensional global ground truth style vector. The codec generator is a decoder primarily based on conformer blocks~\citep{conformer}, similar to MobileSpeech \citep{mobilespeech}. However, we opt for fewer decoder layers (6 layers) and a smaller parameter count in the codec generator.


\section{Evaluation on the out-of-domain style control task.}

\label{appendix outdomain style}

We further evaluate the controllability of style-controllable models with out-of-domain style descriptions. We compare the performance of ControlSpeech with controllable baseline models on the VccmDataset \textbf{\textit{test set C}}. The \textbf{\textit{test set C}} comprises 100 test utterances, \textit{with style prompts rewritten by experts}. None of the test set style prompts are present in the training set.  Results are shown in Table~\ref{table3}. We find that the generalization performance of ControlSpeech is remarkably better than that of the baseline models, which could be attributed to the SMSD module and its underlying mixture density network mechanism. The accuracies of speech speed and volume from ControlSpeech are markedly better than those from baseline models, especially in terms of the volume accuracy. ControlSpeech also yields best WER, MOS-Q, and speaker timbre similarity. Similar to the results shown in Table~\ref{table1}, the pitch accuracy of ControlSpeech is slightly lower. We believe this is due to pitch inconsistencies arising from the simultaneous control of style and timbre cloning. Note that there is no significant difference between the \textbf{\textit{test set A}} and \textbf{\textit{test set C}}, except the style descriptions in \textbf{\textit{test set C}} are out-of-domain while those in \textbf{\textit{test set A}} are in-domain. Comparing Table~\ref{table3} and Table~\ref{table1}, degradations from ControlSpeech on all metrics are much smaller than degradations from baselines.

\begin{table}[htbp]

\caption{The \textbf{out-of-domain style control} results of different style-controlled models on the VccmDataset \textbf{\textit{test set C}}. None of the style prompts are  present in the training set.}
\centering
\begin{adjustbox}{width=0.48\textwidth}
\begin{tabular}{c|cccccc}
\hline
Model  & Pitch $\uparrow$ & Speed $\uparrow$  & Volume $\uparrow$  & WER $\downarrow$ & Spk-sv $\uparrow$ & MOS-Q $\uparrow$ \\
\hline
GT Codec  & 0.85 & 0.87 & 0.91 & 2.8 & 0.96 & 4.25$\pm$0.11 \\ 
Salle  & 0.67 & 0.55 & 0.56  & 6.4 & - & 3.47$\pm$0.08 \\ 
PromptStyle & \textbf{0.77} &0.57 & 0.49& 3.7 & 0.81 & 3.65$\pm$0.11 \\ 
InstructTTS &0.75 & 0.55 & 0.54 & 3.1 & 0.82 & 3.76$\pm$0.14\\ 
PromptTTS 2 & 0.76 & 0.59 &0.58 & 3.3 & - & 3.54$\pm$0.13\\ 
ControlSpeech (\textbf{Ours}) & 0.75 & \textbf{0.73} & \textbf{0.85} & \textbf{3.0} & \textbf{0.88} & \textbf{3.86$\pm$0.12}\\ 
\hline
\end{tabular}
\end{adjustbox}
\label{table3}

\end{table}

\section{Ablation Experiments about Mixed Distributions}
\label{appendix mix}
In this section, we investigate \textbf{the impact of the number of mixtures in the SMSD module on model performance}. We conduct ablation studies under the \textit{isotropic across clusters} noise perturbation mode (the mode selected for ControlSpeech), examining the effects of using 3, 5, and 7 mixtures. As shown in Table~\ref{table5}, the differences in the MOS-SD metric are negligible. However, an increase in the number of mixtures leads to a decline in the MOS-SA metric, indicating that an excessive number of mixtures may reduce the model's control accuracy.

\begin{table}[htbp]
\caption{Under the \textit{Isotropic across clusters} noise perturbation scheme, we investigate the influence of the number of Gaussian mixture components in the SMSD module on stylistic diversity. Subsequently, we analyze the corresponding outcomes using the MOS-SA and MOS-SD metrics.}
\centering
\begin{adjustbox}{width=0.48\textwidth}
\begin{tabular}{c|cc}
\hline
Model & MOS-SA$\uparrow$    & MOS-SD$\uparrow$   \\
\hline
ControlSpeech w/ Isotropic across clusters w/ components=3 & 3.83$\pm$0.14 & 3.95$\pm$0.12\\ 
ControlSpeech w/ Isotropic across clusters w/ components=5 & \textbf{3.84$\pm$0.12} &\textbf{4.05$\pm$0.09} \\ 
ControlSpeech w/ Isotropic across clusters w/ components=7 &3.73$\pm$0.11   &  3.98$\pm$0.09 \\ 
\hline
\end{tabular}
\end{adjustbox}
\label{table5}
\end{table}

\section{Ablation Experiments on Various Noise Modes}
\label{appendix noise}
We analyze the impact of different noise perturbation modes on the many-to-many style control problem, with the number of mixture distributions fixed at 5. As shown in Table~\ref{table6}, we find that the noise perturbation mode maintaining isotropy at the cluster centers achieves a balance between the MOS-SA and MOS-SD metrics and outperforms all other modes.

\begin{table}[htbp]
\caption{The results of different noise perturbation modes on the MOS-SA and MOS-SD metrics.}
\centering
\begin{adjustbox}{width=0.48\textwidth}
\begin{tabular}{c|cc}
\hline
Model & MOS-SA$\uparrow$    & MOS-SD$\uparrow$   \\
\hline
ControlSpeech w/ Fully factored &3.77$\pm$0.14 & 3.96$\pm$0.09\\ 
ControlSpeech w/ Isotropic &3.75$\pm$0.11& 4.03$\pm$0.10\\
ControlSpeech w/ Isotropic across clusters& \textbf{3.84$\pm$0.12} &\textbf{4.05$\pm$0.09} \\ 
ControlSpeech w/ Fixed isotropic&3.72$\pm$0.13   &  3.87$\pm$0.11 \\ 
\hline
\end{tabular}
\end{adjustbox}
\label{table6}
\end{table}

\section{Ethics Statement}
ControlSpeech is capable of zero-shot voice cloning; hence, there are potential risks from misuse, such as voice spoofing. For any real-world applications involving unseen speakers, it is crucial to establish protocols ensuring the speaker's authorization over using the certain speaker's voice. Also, to mitigate these risks, we will also develop approaches such as speech watermarking technology to identify whether a given audio is synthesized by ControlSpeech. 

\section{The SMSD Loss}
\label{appendix smsdloss} 
The loss function for the SMSD module represents the conditional probability of the input style representation ${X_{s}}^{'}$ given the target global style ${Y_{s}}^{'}$. We further refine this into a maximum likelihood loss involving the style distribution parameters $\pi_{k}$, $\mu^{(k)}$, ${\sigma ^{2}}^{(k)}$ derived through the MDN network and noise perturbation module. The detailed derivation of the loss function is as follows.
\onecolumn
\begin{equation}
    \begin{split}
    \mathcal L_{SMSD} & =-logP_{\theta }({Y_{s}}^{'}|{X_{s}}^{'}) \\
    &\propto -\sum_{k=1}^{K}(\pi_{k}exp(-\frac{1}{2}({Y_{s}}^{'}-\mu ^{(k)})^{T}{{\sigma ^{2}}^{(k)}}^{-1}({Y_{s}}^{'}-\mu ^{(k)})-\frac{1}{2}logdet{\sigma ^{2}}^{(k)})) \\
    &= -logsumexp_{k}(log\pi_{k}-\frac{1}{2}({Y_{s}}^{'}-\mu ^{(k)})^{T}{{\sigma ^{2}}^{(k)}}^{-1}({Y_{s}}^{'}-\mu ^{(k)})-\frac{1}{2}logdet{\sigma ^{2}}^{(k)}) \\
    &= -logsumexp_{k}(log\pi_{k}-\frac{1}{2} \left \| \frac{{Y_{s}}^{'}-\mu^{(k)}}{\sigma^{(k)}} \right \|^{2}- \left \| log(\sigma^{(k)}))\right \|_{1})  \\
    &= -logsumexp_{k}(log\pi_{k}-\frac{1}{2} \left \| \frac{{Y_{s}}^{'}-\mu^{(k)}}{\sigma^{(k)}} \right \|^{2}-dlog(\sigma^{(k)}))  \\
    &= -logsumexp_{k}(log\pi_{k}-\frac{1}{2} \left \| \frac{{Y_{s}}^{'}-\mu^{(k)}}{\sigma^{(k)}} \right \|^{2}-dlog(\sigma))  \\
    &=-logsumexp_{k}(log\pi_{k}-\frac{1}{2}\left \| \frac{{Y_{s}}^{'}-\mu^{(k)}}{\sigma }  \right \|^{2} ) \\
    \end{split} 
\end{equation}

\end{document}